\documentclass[preprint,superscriptaddress]{revtex4-1}
\usepackage{cmap}
\usepackage[utf8]{inputenc}
\usepackage[english]{babel}
\usepackage{amssymb,amsfonts,amsmath,mathtext,longtable,array}
\usepackage{xcolor}

\usepackage{graphicx}
\graphicspath{{1/}}

\usepackage[unicode,colorlinks]{hyperref}
\usepackage{breakurl}
\usepackage{soul}
\begin{document}
\UseRawInputEncoding

\title{Non-equilibrium scalar fields at finite temperature and density
}
\author{Sebasti\'an~Mendizabal}
\email{sebastian.mendizabal@usm.cl}

\affiliation{Department of Physics, Universidad T\'ecnica Federico Santa Mar\'ia, Valpara\'iso, Chile.}

\begin{abstract}

We study propagators in bosonic field theories at finite temperature and chemical potential using the Schwinger-Keldysh real-time formalism. The system is considered in contact with a thermal reservoir, allowing for a consistent treatment of both equilibrium and non-equilibrium situations. The chemical potential, associated with conserved charges, modifies the structure of the propagators and introduces features that require detailed analysis.

We focus on how a finite chemical potential affects the analytic structure of the bosonic propagators, including changes in the position of poles and the structure of branch cuts. In our setup, the chemical potential enters the theory as a constant background field, which alters both the dynamics and the boundary conditions. This work provides a basis for understanding the behavior of bosonic fields in thermal and dense environments.
\end{abstract}

%\date{Marz 2025}
\maketitle

%=================================================================================================================
\section{Introduction}
\label{Intro}
%=================================================================================================================
Understanding the behavior of quantum fields at finite temperature and chemical potential is essential for describing a wide range of physical systems, from the early universe and heavy-ion collisions to condensed matter systems with finite density \cite{kolb,Weinberg:2008zzc,lebellac}. In such environments, the standard vacuum quantum field theory framework must be extended to incorporate thermal and statistical effects. Well defined propagators, which encode the causal structure and response of a field, serve as fundamental building blocks for computing physical observables and understanding the microscopic dynamics of interacting many-body systems.
In order to successfully describe out-of-equilibrium phenomena, we will use the Schwinger-Keldysh or closed-time-path (CTP) formalism \cite{Schwinger:1960qe,Keldysh:1964ud,Bakshi:1962dv,Bakshi:1963bn}. Unlike the imaginary-time (Matsubara) formalism \cite{10.1143/PTP.14.351}, which is limited to equilibrium scenarios, the real-time framework allows for a consistent treatment of non-equilibrium dynamics, including time evolution, particle production, and decoherence processes. This makes it particularly suitable for analyzing early-time dynamics, phase transitions, pre-thermalization, and other non-equilibrium phenomena.

In equilibrium, the Kubo-Martin-Schwinger (KMS) condition \cite{Kubo:1957mj,Martin:1959jp} imposes constraints that relate different components of the full propagator in the Keldysh contour, reflecting detailed balance. However, in non-equilibrium scenarios, these relations break down, and one must solve the full set of non-equilibrium equations of motion, the Kadanoff-Baym equations \cite{kadanoff}. These equations can be interpreted as the quantum generalized Boltzmann equations \cite{Drewes:2012qw} and possess all quantum corrections as oppose from the semi-classical Boltzmann ones. We will focus our attention on bosonic Green functions, considering the impact of finite temperature and nonzero chemical potential, specially on the pole structure of the propagators. 

In this work we begin by reviewing the structure of the real-time contour and the resulting formulation of in and out-of-equilibrium propagators, by discussing two physical components, the spectral function and statistical propagator. In equilibrium, this complex systems can be studied by calculating the resulting effective potential, and its convenient expansions at low and high temperatures \cite{Weldon:2007zz,Haber:1981fg,Haber:1981ts,Bernstein:1990kf,Gusynin:2004xr,Benson:1991nj,Loewe:2004zw,Loewe:2005df}. We will discuss our out-of-equilibrium mechanism for these propagators in a strong thermal bath. Later we will focus our attention on the introduction of chemical potential associated to a conserved charge in a simple bosonic scenario. On the other hand we will also focus our attention to the discrepancies" when the chemical potential is introduced in the propagators \cite{Weldon:2007zz,Mallik:2009pj,Sasagawa:2011gn,Sasagawa:2011nh}. One can look at this feature as a new pole structure, or as a new change in the Fourier expansion, with a shifted variable. We will follow the same procedure as \cite{Anisimov:2008dz,Mendizabal:2016zkk} for an out-of-equilibrium scenario with finite chemical potential, where analytical solutions can be obtained for weak chemical potential.

This paper is organized as follows. In Section 2, we review the real-time formalism and define the relevant propagators for a scalar bosonic field, this is essential for understanding the technical features of the problem. In Section 3, we derive the form of the propagators at finite temperature and chemical potential in equilibrium. In Section 4, we extend the analysis to non-equilibrium systems and discuss their time evolution and physical implications. Finally, in Section 5, we summarize our results and outline possible applications.
%=================================================================================================================
\section{Kadanoff-Baym equations (KBE)}
\label{Sec1}
\begin{figure}[htbp]
\centering
\includegraphics[width=\columnwidth]{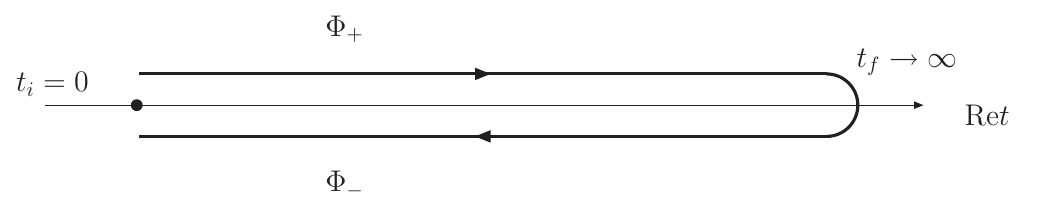}
\caption{Keldysh contour in the complex plane}
\label{fig:keldysh}
\end{figure}

%=================================================================================================================
In quantum field theory, Green's functions are typically defined as the expectation values of time-ordered products of Heisenberg operators evaluated in a given quantum state.
\begin{equation}
    \Delta(t_1,t_2)=-i\Big\langle T\Big(\hat A(t_1)\hat B(t_2)\Big)\Big\rangle\ .
\end{equation}
 By moving to the interaction picture, this expression can be written as:
\begin{equation}
    \Delta(t_1,t_2)=-i\Big\langle T\Big(S^{\dagger}\hat A(t_1)\hat B(t_2)S\Big)\Big\rangle\ ,
\end{equation}
where $S$ is defined by $S\equiv U(\infty,-\infty)=T\exp \Big(-i\int_{-\infty}^{\infty}\mathcal{H}^I_{int}(t)dt\Big).$ For thermal systems, the thermal average of an operator will then be given by $\langle \mathcal{O}\rangle=\text{Tr}(\hat{\rho} \mathcal O)$, where the density matrix operator $\hat \rho =\exp [\beta(\mathcal{F}-\hat{\mathcal{H}})]$ accounts for non pure states. Here, $\beta$ is the inverse of the temperature and $\mathcal F$ is the free energy. By identifying $\beta$ with an imaginary time $it$, the density matrix behaves formally like an evolution operator $\exp⁡[-i\hat Ht]$. To incorporate both thermal and real-time dynamics in the thermal average, we follow the Schwinger-Keldysh formalism \cite{Schwinger:1960qe,Keldysh:1964ud}, in which a complex time contour $\mathcal C$ is introduced, running from $-\infty$ to $\infty$ and back (figure \ref{fig:keldysh}). 
    
%The generalized contour-ordered S-matrix is given by:
%\begin{equation}
%    S_C\equiv T_C\exp \Big(-i\int_C\mathcal{H}^I_{int}(t)dt\Big)\ .
%\end{equation}
%where $T_C$ denotes time ordering along the contour $\mathcal{C}$. Correspondingly
%The contour-ordered Green’s function is defined as:
%\begin{equation}
%    \Delta_C(t_1,t_2)=-i\Big \langle T_C(\hat A(t_1)\hat B(t_2)\Big\rangle\ .
%\end{equation}
For a scalar field $\Phi(x)$ in the $\mathcal{C}$-contour we have the Green function given by:
\begin{equation}\label{deltadefinition}
    \Delta_C(x_1,x_2)=\theta(x^0_1,x^0_2)\Delta^>(x_1,x_2)+\theta(x^0_2,x^0_1)\Delta^<(x_1,x_2)\ ,
\end{equation}
 where 
 \begin{eqnarray}
      \Delta^>(x_1,x_2)&\equiv&\langle\Phi(x_1)\Phi(x_2)\rangle \ , \\
      \Delta^<(x_1,x_2)&\equiv&\langle\Phi(x_2)\Phi(x_1)\rangle \ .
 \end{eqnarray}
One can write the Schwinger-Dyson equation in the contour:
 \begin{equation}\label{schwingerdysoneq}
     (\square_1+m^2)\Delta_C(x_1,x_2)+i\int_Cd^4x'\Pi_C(x_1,x')\Delta_C(x',x_2)=-i\delta_C(x_1-x_2)\ ,
 \end{equation}
 where $\Pi_C$ is defined in the same way as in (\ref{deltadefinition}), and $\square_1=\partial^2/\partial t_1^2-\nabla^2$. One can also define the symmetric and antisymmetric correlators, fort the Green functions and self energies
 \begin{eqnarray}\label{plusminusdefinitiosa}
     \mathcal O^+(x_1,x_2)\equiv & \frac{1}{2}\big( \mathcal O^>(x_1,x_2)+\mathcal O^<(x_1,x_2)\big)\ ,\\\label{plusminusdefinitiosb}
     \mathcal O^-(x_1,x_2)\equiv &  i\big( \mathcal O^>(x_1,x_2)-\mathcal O^<(x_1,x_2)\big)\ .
 \end{eqnarray}
% which will define the retarded and advance correlators
 %\begin{eqnarray}
%      \mathcal O^R(x_1,x_2)\equiv & i\theta(t_1-t_2)( \mathcal O^>(x_1,x_2)-\mathcal %O^<(x_1,x_2))=\theta(t_1-t_2)\mathcal O^-(x_1,x_2)\ ,\\
  %    \mathcal O^A(x_1,x_2)\equiv & -i\theta(t_2-t_1)( \mathcal O^>(x_1,x_2)-\mathcal %O^<(x_1,x_2))=-\theta(t_1-t_2)\mathcal O^-(x_1,x_2)\ .
 %\end{eqnarray}

With these definitions, the Schwinger-Dyson equation (\ref{schwingerdysoneq}) in a spatial translational invariant system, can be written with two coupled differential equations
\begin{eqnarray}
    \big(\partial^2_{t_1}& +&\omega_{\bf k}^2\big)\Delta_{\bf k}^-(t_1,t_2)+\int_{t_2}^{t_1}dt'\Pi_{\bf k}^-(t_1,t')\Delta_{\bf k}^-(t',t_2)=0\ ,\\
    \big(\partial^2_{t_1}& +&\omega_{\bf k}^2\big)\Delta_{\bf k}^+(t_1,t_2)+\int_{t_i}^{t_1}dt'\Pi_{\bf k}^-(t_1,t')\Delta_{\bf k}^+(t',t_2)=\int_{t_i}^{t_2}dt'\Pi_{\bf k}^+(t_1,t')\Delta_{\bf k}^-(t',t_2)\ ,
\end{eqnarray}
where $\partial_{t_1}=\partial/\partial t_1$ and $\omega_{\bf k}^2=k^2+m^2$. It was shown in \cite{Anisimov:2008dz} that for a time translational invariant background, where:
\begin{equation}\label{eqselenergy}
    \Pi_{\bf k}(t_1,t_2)=\Pi_{\bf k}(t_1-t_2)\ ,
\end{equation}
the spectral function con also be expressed as a function of the difference of the time variables:
\begin{equation}\label{eqspectralfunction}
    \Delta^-_{\bf k}(t_1,t_2)=\Delta^-_{\bf k}(t_1-t_2)\ .
\end{equation}
Finally, we obtain a set of differential-integral equations
\begin{eqnarray}\label{KBequations}
    (\partial_{t_1}^2+\omega_{\bf k}^2)\Delta^-_{\bf k}(t_1-t_2)=&&-\int_{t_1}^{t_2}dt'\Pi^-_{\bf k}(t_1-t')\Delta_{\bf k}^-(t'-t_2)\ ,\\ \label{secondKBequation}
    (\partial_{t_1}^2+\omega_{\bf k}^2)\Delta^+_{\bf k}(t_1,t_2)=&&- \int_{t_i}^{t_1}dt'\Pi^-_{\bf k}(t_1-t')\Delta^+_{\bf k}(t',t_2)\nonumber \\&&+\int_{t_i}^{t_2}dt'\Pi^+_{\bf k}(t_1-t')\Delta^-_{\bf k}(t'-t_2)\ .
\end{eqnarray}
In this scenario, analytical solutions can be obtained for both the spectral function ($\Delta^-$) and the statistical propagator ($\Delta^+$). Under equilibrium conditions, these propagators are constrained by the Kubo-Martin-Schwinger (KMS) condition, which relates equilibrium correlation functions. In frequency space ($\omega$-space), this relation takes the form:
 \begin{equation}\label{kms+-}
     \Delta_{\bf k}^<(\omega)=e^{-\beta\omega}\Delta_{\bf k}^>(\omega)\ .
 \end{equation}
This leads to the KMS relation for the statistical and spectral propagators, $\Delta_{\bf k}^+(\omega)$ and $\Delta_{\bf k}^-(\omega)$, respectively:
 \begin{equation}\label{kmsmasymenos}
     \Delta^+_{\bf k}(\omega)=-\frac{i}{2}\coth{\Big(\frac{\beta \omega}{2}\Big)}\Delta^-_{\bf k}(\omega)\ .
 \end{equation}
Note that, in equilibrium, the statistical propagator depends only on the time difference, i.e., $\Delta^+_{\bf k}(t_1,t_2)=\Delta^+_{\bf k}(t_1-t_2)$.
 
 \section*{Solution to the first Kadanoff-Baym equations}
 
Solutions to the Kadanoff-Baym equations in (\ref{KBequations}) and (\ref{secondKBequation}) can be obtain analytically just by considering an equilibrium reservoir in which (\ref{eqselenergy}) and (\ref{eqspectralfunction}) are satisfied to first order, neglecting, for example, backreaction. Systems where back-reaction are strong enough to be relevant are discussed in \cite{Mendizabal:2016zkk}. We begin by focusing on the first, homogeneous Kadanoff–Baym equation for the spectral function $\Delta^-_{\bf k}(y)$, where we define $y\equiv t_1-t_2$:
 \begin{equation}\label{KB1}
     (\partial_{y}^2+\omega_{\bf k})\Delta^-_{\bf k}(y)+\int_{0}^{y}dy'\Pi^-_{\bf k}(y-y')\Delta_{\bf k}(y')=0\ .
 \end{equation}
 The initial conditions can be determined from the symmetry properties of the correlator, namely:
 \begin{equation}
     \Delta_{\bf k}^-(y)|_{y=0}=0\ ,\hspace{1cm}
     \partial_y\Delta_{\bf k}^-(y)|_{y=0}=1\ .
 \end{equation}
 The solution to equation (\ref{KB1}) can then be obtained by performing a Laplace transform with respect to the variable $y$:
 \begin{equation}
     \Delta_{\bf k}^-(y)=\int_{C_B}\frac{ds}{2\pi i}\frac{e^{sy}}{s^2+\omega_{\bf k}^2+\tilde {\Pi}_{\bf k}^-(\omega)}\ .
 \end{equation}
Here, \( C_B \) denotes the Bromwich contour used in the inverse Laplace transformation. The solution can be obtained by employing the relation between the self-energies,
\[
\tilde{\Pi}_{\bf k}^-(i\omega \pm \epsilon) = \text{Re}\, \Pi_{\bf k}^R(\omega) \pm i\, \text{Im}\, \Pi_{\bf k}^R(\omega)\ ,
\]
where the imaginary part of the retarded self-energy is given by
\[
\text{Im}\, \Pi_{\bf k}^R(\omega) = \frac{1}{2i} \left[ \Pi_{\bf k}^R(\omega + i\epsilon) - \Pi_{\bf k}^R(\omega - i\epsilon) \right]\ .
\]
This leads to the following expression:
\begin{equation}\label{spectraltimedependance}
\Delta_{\bf k}^-(y) = i \int_{-\infty}^{\infty} \frac{d\omega}{2\pi} \, e^{-i\omega y} \, \frac{-2\, \text{Im}\, \Pi_{\bf k}^R(\omega) + 2\omega \epsilon}{\left[\omega^2 - \omega_{\bf k}^2 - \text{Re}\, \Pi_{\bf k}^R(\omega)\right]^2 + \left[\text{Im}\, \Pi_{\bf k}^R(\omega) + \omega \epsilon\right]^2}\ .
\end{equation}

A quasiparticle interpretation can be introduced by identifying the dominant poles of the integrand, leading to the definitions:
\begin{equation}\label{realpolescalar}
\Omega_{\bf k} \cong \omega_{\bf k} + \frac{\text{Re}\, \Pi_{\bf k}(\omega_{\bf k})}{2\omega_{\bf k}}\ , \hspace{1cm} \Gamma_{\bf k} \cong -\frac{1}{\Omega_{\bf k}} \text{Im}\, \Pi_{\bf k}^R(\Omega_{\bf k})\ .
\end{equation}

Substituting equations (\ref{realpolescalar}) into (\ref{spectraltimedependance}) and taking the limit \( \epsilon \rightarrow 0 \), the integral yields:
\begin{equation}\label{spectraltimedepenence}
\Delta_{\bf k}^-(y) = \frac{1}{\Omega_{\bf k}} \sin(\Omega_{\bf k} y) \, e^{-\frac{\Gamma_{\bf k}}{2} |y|}\ .
\end{equation}

Equations (\ref{spectraltimedepenence}) and (\ref{spectraltimedependance}) represent the standard Breit-Wigner form of the equilibrium spectral function. This result will be employed in the computation of the statistical propagator.
 %%%%%%%%%%%%%%%%%%%
 \section*{Solution to the second Kadanoff-Baym equations}
The second KB equation is given by:
\begin{eqnarray}
    (\partial_{t_1}^2+\omega_{\bf k}^2)\Delta^+_{\bf k}(t_1,t_2)=&&- \int_{t_i}^{t_1}dt'\Pi^-_{\bf k}(t_1-t')\Delta^+_{\bf k}(t',t_2)\nonumber \\&&+\int_{t_i}^{t_2}dt'\Pi^+_{\bf k}(t_1-t')\Delta^-_{\bf k}(t'-t_2)\ .
  \end{eqnarray}
This is a highly nontrivial integro-differential equation, but it admits an analytical solution within a grand-canonical framework, where the reservoir exhibits time-translational invariance, as indicated in equation (\ref{eqselenergy}). We will now explicitly detail the steps of this calculation, as they will serve as a reference for the analogous procedure when a nonzero chemical potential is introduced.
In this scenario, the solution can be separated and expressed as follows:
 \begin{equation}
 \Delta_{\bf k}^+(t_1,t_2)=\hat{\Delta}_{\bf k}^+(t_1,t_2)+\Delta_{\bf k}^{mem}(t_1,t_2)\ ,
 \end{equation}
 where $\hat{\Delta}_{\bf k}^+(t_1,t_2)$ is an homogeneous solution, while the second term corresponds to the inhomogeneous one and is given by:
 \begin{equation}
     \Delta_{\bf k}^{mem}(t_1,t_2)=\int_{t_i}^{t_1}dt'\int_{t_i}^{t_2}dt''\Delta_{\bf k}^-(t_1-t')\Pi_{\bf k}^+(t'-t'')\Delta_{\bf k}^-(t''-t_2)\ .
 \end{equation}
Let us begin with the inhomogeneous solution. To properly integrate it, we express it as follows:
  \begin{equation}
     \Delta_{\bf k}^{mem}(t_1,t_2)=\int_{t_i}^{t_1}dt'\int_{t_i}^{t_2}dt''\Delta_{\bf k}^-(t_1-t')\bigg(\int\frac{d\omega}{2\pi}e^{-i\omega(t'-t'')}\Pi_{\bf k}^+(\omega)\bigg)\Delta_{\bf k}^-(t''-t_2)\ ,
 \end{equation}
 so we can separate the time integrals
   \begin{eqnarray}
     \Delta_{\bf k}^{mem}(t_1,t_2)=&&\int\frac{d\omega}{2\pi}\bigg(\int_{t_i}^{t_1}dt'\Delta_{\bf k}^-(t_1-t')e^{i\omega(t_1-t')}\bigg)\Pi_{\bf k}^+(\omega)\nonumber\\
     &&\times\bigg(\int_{t_i}^{t_2}dt''\Delta_{\bf k}^-(t''-t_2)e^{-i\omega(t_2-t'')}\bigg)e^{-i\omega(t_1-t_2)}\ .
 \end{eqnarray}
 Taking $t_i=0$ and defining $y=t_1-t_2$, $y_1=t_1-t'$ and $y_2=t_2-t''$ we obtain
 \begin{equation}\label{statiticalproptime}
     \Delta_{\bf k}^{mem}(t_1,t_2)=\int\frac{d\omega}{2\pi}\bigg(\int_{0}^{t_1}dy_1\Delta_{\bf k}^-(y_1)e^{i\omega y_1}\bigg)\Pi_{\bf k}^+(\omega)\bigg(\int_{0}^{t_2}dy_2\Delta_{\bf k}^-(-y_2)e^{-i\omega y_2}\bigg)e^{-i\omega y}\ .
 \end{equation}
The procedure proceeds as follows: First, we integrate over the time variables \( y_1 \) and \( y_2 \), then identify the pole structure. This allows us to define the quasiparticle’s new frequency \( \Omega_{\bf k} \) and decay parameter \( \Gamma_{\bf k} \), which depend on the real and imaginary parts of the self-energy. We then perform the integration over the frequency \( \omega \) to obtain the inhomogeneous contribution. It is important to note that any change in the spectral function \( \Delta^-_{\bf k} \) will be reflected in equation (\ref{statiticalproptime}) for the statistical propagator. The decay parameter, as defined in equation (\ref{realpolescalar}), will be used after evaluating the poles following the \( \omega \)-integration. To proceed with this integral, it is convenient to rewrite equation (\ref{statiticalproptime}).
 \begin{equation}\label{statiticalproptime}
     \Delta_{\bf k}^{mem}(t_1,t_2)=\int\frac{d\omega}{2\pi}\mathcal I(t_1;\omega)\Pi_{\bf k}^+(\omega)\mathcal I(t_2;\omega)e^{-i\omega y}\ ,
 \end{equation}
where
\begin{equation}\label{statiticalproptimeconies}
    \mathcal I(t_j;\omega)=\int_{0}^{t_j}dy'\Delta_{\bf k}^-(\pm y')e^{\pm i\omega y'}\ .
\end{equation}
Here, $t_j=t_1t_2$, and the ±± sign will be applied to $t_1$ and $t_2$ correspondingly. Note that we can drop the absolute value in the spectral propagator, as the integral is defined between $0$ and $t$. The equilibrium scalar propagator, obtained from the first Kadanoff-Baym equation in equation (\ref{spectraltimedepenence}), can then be substituted into equation (\ref{statiticalproptimeconies}) to yield:
\begin{equation}
    \mathcal I(t_1,\Omega)=\frac{-1}{\Omega_{\bf k}(\Omega_{\bf k}^2-(\Omega+i\Gamma_{\bf k})^2)^2}\bigg[(i\omega\sin(\Omega_{\bf k} t_1)-\Omega_{\bf k}\cos(\Omega_{\bf k} t_1))e^{i(\omega+i\Gamma_{\bf k}/2)t_1}+\Omega_{\bf k}\bigg]\ ,
\end{equation}
and
\begin{equation}
    \mathcal I(t_2,\Omega)=\frac{-1}{\Omega_{\bf k}(\Omega_{\bf k}^2-(\Omega-i\Gamma_{\bf k})^2)^2}\bigg[(i\omega\sin(\Omega_{\bf k} t_2)+\Omega_{\bf k}\cos(\Omega_{\bf k} t_2))e^{i(\omega-i\Gamma_{\bf k}/2)t_2}-\Omega_{\bf k}\bigg]\ .
\end{equation}
Since the reservoir is in the thermal equilibrium, the self-energy $\Pi_{\bf k}^+(\omega)$ can be computed using the KMS condition in (\ref{kms+-}). Additionally, given that $\Pi_{\bf k}^-(\omega)=2i\text{Im}\Pi_{\bf k}^R(\omega)$, we can express the symmetric self-energy in terms of the imaginary part of the retarded self-energy as follows:
\begin{equation}
\Pi_{\bf k}^+(\omega)=\coth\bigg(\frac{\beta \omega}{3}\bigg)\text{Im}\Pi_{\bf k}^R(\omega)\ .
\end{equation}
This will give us 
\begin{eqnarray}
    \Delta_{\bf k}^{mem}(t_1,t_2)=&-&\frac{i}{2\Omega_{\bf k}^2}\int\frac{d\omega}{2\pi}\frac{e^{-i\omega(t_1-t_2)}\coth\big(\frac{\beta\omega}{2}\big)\text{Im}\Pi_{\bf k}^R(\omega)}{\big(\Omega_{\bf k}^2-(\omega+i\Gamma_{\bf k})^2\big)\big(\Omega_{\bf k}^2-(\omega-i\Gamma_{\bf k})^2\big)}\nonumber\\
    &\times&\Big[(i\omega\sin(\Omega_{\bf k} t_1)-\Omega_{\bf k}\cos(\Omega_{\bf k} t_1))e^{i(\omega+i\Gamma_{\bf k}/2)t_1}+\Omega_{\bf k}\Big]\nonumber\\
    &\times&\Big[(i\omega\sin(\Omega_{\bf k} t_2)+\Omega_{\bf k}\cos(\Omega_{\bf k} t_2))e^{i(\omega-i\Gamma_{\bf k}/2)t_2}-\Omega_{\bf k}\Big ]\ .
\end{eqnarray}
The poles of this equation are located (up to first order in $\Gamma_{\bf k}$) at: $\omega=\pm\Omega_{\bf k}\pm i\Gamma_{\bf k}/2$. Each term must be integrated separately to ensure convergence. Following the approach in \cite{Anisimov:2008dz}, we define $S_i=\sin(\Omega_{\bf k})e^{i(\omega+i\Gamma_{\bf k}/2)t_i}$, and $C_i=\cos(\Omega_{\bf k})e^{i(\omega+i\Gamma_{\bf k}/2)t_i}$. This allows us to write:
\begin{eqnarray}
        \Delta_{\bf k}^{mem}(t_1,t_2)=&-&\frac{i}{2\Omega_{\bf k}^3}\int\frac{d\omega}{2\pi}\frac{e^{-i\omega(t_1-t_2)}\coth\big(\frac{\beta\omega}{2}\big)\text{Im}\Pi_{\bf k}^R(\omega)}{|\Omega_{\bf k}^2-(\omega+i\Gamma_{\bf k})^2|^2}\nonumber\\
    &\times&\Big[\Omega_{\bf k}^2\big(-S_1S_2^*-C_1C_2^*+C_1+C_2^*-1\big)\nonumber\\
    &&+i\Omega_{\bf k}\omega\big(S_1C_2^*-C_1S_2^*-S_1+S_2^*\big)\Big ]\ .
\end{eqnarray}
After evaluating at the poles, we can make a change in the numerator, up to first order in $\Gamma_{\bf k}$.
\begin{eqnarray}
    \omega&\rightarrow&\Omega_{\bf k}\text{Sign}(\omega)\ ,\\
    \coth\Big(\frac{\beta\omega}{2}\Big)&\rightarrow& \coth\Big(\frac{\beta\Omega_{\bf k}}{2}\Big)\text{Sign}(\omega)\ ,\\
    \text{Im}\Pi_{\bf k}^R(\omega)&\rightarrow&-\Omega_{\bf k}\Gamma_{\bf k}\text{Sign}(\omega)\ .
\end{eqnarray}
On the other hand, the denominator, after integration will always have a factor $\pm(4i\Omega_{\bf k}\Gamma_{\bf k})^{-1},$
where $\pm$ is related to closing the path anticlockwise in the upper path, and clockwise on the lower path respectively. After a lengthy calculations it is straightforward to obtain
\begin{equation}
      \Delta_{\bf k}^{mem}(t_1,t_2)=\frac{1}{\Omega_{\bf k}}\cos\big(\Omega_{\bf k}(t_1-t_2)\big)\coth\Big(\frac{\beta\Omega_{\bf k}}{2}\Big)\bigg[e^{-\frac{\Gamma_{\bf k}}{2}(t_1+t_2)}-e^{-\frac{\Gamma_{\bf k}}{2}|t_1-t_2|}\bigg]\ .
\end{equation}
This solution shows the non-equilibrium effects associated with the decay parameter. As expected, this term vanishes for large times \( t = \frac{t_1 + t_2}{2} \). Notice the use of the absolute value for the time difference in the second exponential, which arises from the integration over the poles. This ensures that \( \Delta_{\bf k}^+(t_1,t_2) = \Delta_{\bf k}^+(t_2,t_1) \). The solution for the homogeneous part can be found by following the approach in \cite{Anisimov:2008dz}, where:
\begin{eqnarray}\label{initialcoditionsdeltaplus}
    \hat{\Delta}_{\bf k}^+(t_1,t_2)=&&\Delta_{in}^+\dot{\Delta}_{\bf k}^-(t_1)\dot{\Delta}_{\bf k}^-(t_2)+\ddot{\Delta}_{in}^+\Delta_{\bf k}^-(t_1)\Delta_{\bf k}^-(t_2)\nonumber\\
    &+&\dot{\Delta}_{in}^+\big(\dot{\Delta}_{\bf k}^-(t_1)\Delta_{\bf k}^-(t_2)+\Delta_{\bf k}^-(t_1)\dot{\Delta}_{\bf k}^-(t_2)\big)\ .
\end{eqnarray}
For initial conditions given by
\begin{equation}
   \Delta_{in}=\Omega_{\bf k};\hspace{0.5cm} \dot{\Delta}_{in}=0;\hspace{0.5cm}\dot{\Delta}_{in}=\Omega_{\bf k},
\end{equation}
we finally get the full non-equilibrium scalar propagator 
\begin{equation}
    \Delta_{\bf k}^+(t_1,t_2)=\frac{1}{\Omega_{\bf k}}\cos\big(\Omega_{\bf k}(t_1-t_2)\big)\bigg[\frac{1}{2}\coth\bigg(\frac{\beta\Omega_{\bf k}}{2}\bigg)e^{-\frac{\Gamma_{\bf k}}{2}|t_1-t_2|}+f_B^{eq}(\Omega_{\bf k})e^{\frac{-\Gamma_{\bf k}}{2}(t_1+t_2)}\bigg]\ ,
\end{equation}
where $f_B^{eq}(\Omega_{\bf k})=(e^{\beta\Omega_{\bf k}}-1)^{-1}$ is the equilibrium distribution function of a scalar field with energy $\Omega_{\bf k}$. Using the definitions in equations (\ref{plusminusdefinitiosa}) and (\ref{plusminusdefinitiosb}), one can derive the full list of non-equilibrium propagators. We will use this result as a basis for comparison with the full propagator that includes a non-vanishing chemical potential.

%=================================================================================================================
\section{Equilibrium propagators at finite chemical potential}
\label{s2}
%=================================================================================================================
The introduction of a chemical potential associated with a global symmetry is typically implemented via the density operator:
\begin{equation}
    \rho=e^{-\beta(\hat H-\mu\hat Q)}\ ,
\end{equation}
where $\hat Q$ denotes the conserved charge operator associated with the symmetry  \cite{Weldon:2007zz,Haber:1981fg,Haber:1981ts,Gusynin:2004xr,Bernstein:1990kf,Benson:1991nj}. Alternatively, the chemical potential can be introduced via a constant background field  \cite{Loewe:2004zw,Loewe:2005df}, by modifying the derivative as $\partial_{\mu}\rightarrow \partial_{\mu}+i \mu\delta^{\mu}_{0}$. There has been considerable discussion regarding whether the chemical potential should be incorporated directly into the action or instead introduced through modifications to the Kubo-Martin-Schwinger (KMS) condition. Both approaches are formally equivalent and lead to the same physical results. Nonetheless, care must be taken when performing variable transformations and Fourier analysis, as the presence of a chemical potential affects these operations as well as the symmetry properties of the correlation functions.

In order to compute non-equilibrium propagators in the presence of a chemical potential, it must be consistently included in the Kadanoff–Baym equations, following the procedure outlined in Section II. Adopting the conventions of \textbackslash{}cite\{weldon\}, one can derive the first Kadanoff–Baym equation for the spectral function starting from an action that explicitly includes the chemical potential:
\begin{equation}\label{KBdeltaminuschemical}
    \big((\partial_0-i\mu)^2+\omega_{\bf k}^2\big)G_{\mu,\bf k}^{-}(y)+\int_0^ydy'\Pi_{\bf k}^-(y-y')G_{\mu,\bf k}^-(y')=0\ ,
\end{equation}
where $G_{\mu,\bf k}^{-}(y)$ is a modified spectral function with the presence of a chemical potential. Since the chemical potential enters as a constant background in the time component of the derivative, the KMS conditions given in equations (\ref{kms+-}) and (\ref{kmsmasymenos}) continue to apply to $\Delta_{\mu,\bf k}^{-}(y)$. The procedure for computing the spectral function remains unchanged from the case without a chemical potential. By performing a Laplace transformation, one can derive from equation  (\ref{KBdeltaminuschemical}) that
\begin{equation}
    \tilde{G}_{\mu,\bf k}^-(s)=\frac{\partial_0 G_{\mu,\bf k}^-(0)+sG_{\mu,\bf k}^-(0)-2i\mu G_{\mu,\bf k}^-(0)}{s^2-2i\mu s+(\omega_{\bf k}^2-\mu^2)+\tilde{\Pi}_{\bf k}(s)}\ ,
\end{equation}
where, as before, $G_{\mu,\bf k}^-(0)$ and $\partial_0G_{\mu,\bf k}^-(0)$ can be obtain from symmetry properties. The inverse of this equation can be found integrating over the Bromwich contour.
\begin{equation}\label{deltaminusbromwich}
G_{\mu,\bf k}^-(y)=
\int_{\mathcal C_B}\frac{ds}{2\pi i}\frac{e^{sy}\big(\partial_0G_{\mu,\bf k}^-(0)+sG_{\mu,\bf k}^-(0)-2i\mu G_{\mu,\bf k}^-(0)\big)}{s^2-2i\mu s+(\omega_{\bf k}^2-\mu^2)+\tilde{\Pi}_{\bf k}(s)}\ .
\end{equation}
At this stage, two approaches can be pursued: one may either retain the chemical potential explicitly in the denominator or perform a variable shift  $x\rightarrow x-\mu$, where $x$ is related to the integration variable. In the case of vanishing self-energy and a finite chemical potential, the solution to the equation above takes the form:
\begin{equation}
    G_{\mu,\bf k}^-(y)=\frac{1}{\omega_{\bf k}}\sin(\omega_{\bf k}y)e^{i\mu y}\ ,
\end{equation}
or in the $\omega$-space 
\begin{equation}\label{dsdsds}
    G_{\mu,\bf k}^-(\omega)=\frac{4\epsilon(\omega-\mu)}{[(\omega-\mu)^2-\omega_{\bf k}^2]^2+4\epsilon^2(\omega - \mu )^2}\ ,
\end{equation}
with $\epsilon\rightarrow 0$ acting as a regulator. For the $G_{\mu,\bf k}^>(y)$ and $G_{\mu,\bf k}^<(y)$ propagators we get:
\begin{equation}
    G_{\mu,\bf k}^{<}(y)=\frac{e^{i\mu y}}{2\omega_{\bf k}}\bigg[e^{i\omega_{\bf k}}f^{eq}_B(\omega_{\bf k}-\mu)+e^{-i\omega_{\bf k} y}[1+f_B^{eq}(\omega_{\bf k}+\mu)]\bigg]\ ,
\end{equation}
and
\begin{equation}
    G_{\mu,\bf k}^{<}(y)=\frac{e^{i\mu y}}{2\omega_{\bf k}}\bigg[e^{i\omega_{\bf k}}[1+f_B^{eq}(\omega_{\bf k}+\mu)]+e^{-i\omega_{\bf k} y}f^{eq}_B(\omega_{\bf k}-\mu)\bigg]\ .
\end{equation}
Finally, for the equilibrium statistical propagator with vanishing self-energy but finite chemical potential we obtain:
\begin{equation}
    G_{\mu,\bf k}^+(y)=\frac{1}{2\omega_{\bf k}}\cos(\omega_{\bf k} y)e^{i\mu y}\big[1+f_B^{eq}(\omega_{\bf k}-\mu)+f_B^{eq}(\omega_{\bf k}+\mu)\big]\ .
\end{equation}
The statistical propagator in the $\omega$-space can be obtain using the KMS condition in (\ref{kmsmasymenos}).

For the case of a non-vanishing self-energy, the calculation are straightforward following eq. (\ref{deltaminusbromwich}), or the analogue to equation (\ref{dsdsds})
\begin{equation}\label{gimutiempoint}
    G_{\mu,\bf k}^-(y)=i\int\frac{d\omega}{2\pi}e^{i\omega y}e^{i\mu y} \rho_{\mu,\bf k}(\omega)\ ,
\end{equation}
with
\begin{equation}\label{dsdsds}
    \rho_{\mu,\bf k}(\omega)=\frac{2\text{Im}\Pi^R(\omega+\mu)+4\epsilon\omega}{[\omega^2-\omega_{\bf k}^2-\text{Re}\Pi_{\bf k}^R(\omega+\mu)]^2+[\text{Im}\Pi^R(\omega+\mu)+2\epsilon\omega ]^2}\ .
\end{equation}
The way the spectral function is written in the above equations shows that the chemical potential is related to a constant in the Fourier transformation on the variable $y$. It follows that for a transformation  $y\rightarrow -y$, one should also take $\mu\rightarrow -\mu$. 

The procedure for integrating equation (\ref{gimutiempoint}) is the same as before, by defining 
\begin{equation}
    \Omega_{\mu,\bf k}^{\pm 2}=\omega_{\bf k}^2+\text{Re}\Pi_{\bf k}^R(\omega_{\bf k}\pm\mu)\ ,
\end{equation}
and
\begin{equation}
    \Gamma_{\mu,\bf k}^{\pm}(\omega)=\frac{\text{Im}\Pi_{\mu,\bf k}^R(\Omega_{\mu,\bf k}^{\pm}\pm\mu)}{\Omega_{\mu,\bf k} ^{\pm}}\text{Sign}(\Omega_{\bf k}^{\pm})\ ,
\end{equation}
By analyzing the pole structure of the spectral function, one finds a mixture of distinct pole contributions. Incorporating the $\text{Sign}$ function into the modified decay parameters leads to a new expression for the spectral function:
\begin{equation}\label{spectralfunctminusplus}
     G^-_{\bf k}(y)=\frac{e^{i\mu y}}{2i}\bigg[\frac{e^{i\Omega_{\bf k}^+y}e^{-\Gamma_{\mu,\bf k}^+|y|/2}}{\Omega^+_{\bf k}}-\frac{e^{-i\Omega_{\bf k}^-y}e^{-\Gamma_{\mu,\bf k}^-|y|/2}}{\Omega^-_{\bf k}}\bigg]\ .
 \end{equation}
 The fact that the sign of the chemical potential separates the pole structure defined in $\Omega_{\bf k}^{\pm}$, with all the positive signs separated from the minus signs is not a coincidence, and will have big impact in the statistical propagator. Notice that for small chemical potential, one regains the result:
 \begin{equation}\label{spectralfunctminusplusapprox}
     G^-_{\bf k}(y)=\frac{e^{i\mu y}}{\Omega_{k}}\sin\big(\Omega_{\bf k}\big)e^{-\Gamma_{\bf k}|y|/2}\ ,
 \end{equation}
with the chemical potential acting as a phase. The statistical propagator can be found using the KMS condition.

%=================================================================================================================
\section{Non-equilibrium propagators at finite chemical potential}
\label{s3}
%=================================================================================================================
In the case of vanishing chemical potential, we restrict our analysis to an equilibrium thermal background, ensuring that no back-reaction affects the spectral function. As a result, both the self-energies and the spectral function depend only on the time difference, while the statistical propagator retains a dependence on the mean time, defined as ($t=(t_1+t_2)/2$). For big chemical potential the spectral function is given by (\ref{spectralfunctminusplus}), and for small chemical potential by (\ref{spectralfunctminusplusapprox}), where the approximation
\begin{equation}
    \Omega_{\bf k}^{\pm}\rightarrow \Omega_{\bf k},\hspace{0.3 cm} \text{and}\hspace{0.3cm}  \Gamma_{\bf k}^{\pm}\rightarrow \Gamma_{\bf k}
\end{equation}
can be used. For a complete treatment of the statistical propagator, one must formulate the second Kadanoff–Baym equation (\ref{secondKBequation}), now extended to include a finite chemical potential.
\begin{eqnarray}
\label{secondKBequationfinitemu}
    \big((\partial_{t_1}-i\mu)^2+\omega_{\bf k}^2\big)G^+_{\mu,\bf k}(t_1,t_2)=&&\int_{t_i}^{t_2}dt'\Pi^+_{\bf k}(t_1-t')G^-_{\mu,\bf k}(t'-t_2)\nonumber \\
   -&& \int_{t_i}^{t_1}dt'\Pi^-_{\bf k}(t_1-t')G^+_{\mu,\bf k}(t',t_2)\ .
\end{eqnarray}
As usual, the solution for is a sum of an homogeneous and an inhomogeneous one, i.e., $G_{\mu,\bf k}^+(t_1,t_2)=\hat{G}_{\mu,\bf k}^{+}(t_1,t_2)+G_{\mu,\bf k}^{mem}(t_1,t_2)$. We start with the calculation of the inhomogeneous solution $G_{\mu,\bf k}^{mem}(t_1,t_2)$, we can write it as 
\begin{eqnarray}\label{gmemnonequichem}
     G_{\mu,\bf k}^{mem}(t_1,t_2)=&&\int\frac{d\omega}{2\pi}\bigg(\int_{t_i}^{t_1}dt'G_{\mu,\bf k}^-(t_1-t')e^{i\omega(t_1-t')}\bigg)\Pi_{\bf k}^+(\omega)\nonumber\\ &\times&\bigg(\int_{t_i}^{t_2}dt''G_{\mu,\bf k}^-(t''-t_2)e^{-i\omega(t_2-t'')}\bigg)e^{-i\omega(t_1-t_2)}\ ,
 \end{eqnarray}
Without loss of generality, we can set $t_i=0$. In this formulation, the chemical potential enters through the solution of the spectral function. As in previous cases, the time integration can be performed independently, resulting in a modified pole structure. This, in turn, affects the sign of the chemical potential appearing in the exponential terms as well as in the self-energies. We can then define:
\begin{equation}
    I_1(\omega)=\int_{0}^{t_1}dt'G_{\mu,\bf k}^-(t_1-t')e^{i\omega(t_1-t')}\ ,
\end{equation}
and
\begin{equation}
    I_2(\omega)=\int_{0}^{t_2}dt''G_{\mu,\bf k}^-(t''-t_2)e^{-i\omega(t_2-t'')}\ .
\end{equation}
This two integrals will give 
\begin{eqnarray}\label{i1chemicalnonequi}
    I_1(\omega)=-\frac{1}{2}\Bigg[\frac{e^{i(\Omega_{\mu,\bf k}^++\mu)t_1}}{\Omega_{\mu,\bf k}^+}\frac{e^{i(\omega+i\frac{\Gamma_{\mu,\bf k}^+}{2})t_1}-1}{\big(\omega+\Omega_{\mu,\bf k}^++\mu+\frac{i\Gamma_{\mu,bf k}^+}{2}\big)}-\frac{e^{i(-\Omega_{\mu,\bf k}^-+\mu)t_1}}{\Omega_{\mu,\bf k}^-}\frac{e^{i(\omega+i\frac{\Gamma_{\mu,\bf k}^-}{2})t_1}+1}{\big(\omega-\Omega_{\mu,\bf k}^-+\mu+\frac{i\Gamma_{\mu,\bf k}^-}{2}\big)}\Bigg]\ ,
\end{eqnarray}
and
\begin{eqnarray}\label{i2chemicalnonequi}
    I_2(\omega)=-\frac{1}{2}\Bigg[\frac{e^{i(-\Omega_{\mu,\bf k}^+-\mu)t_2}}{\Omega_{\mu,\bf k}^+}\frac{e^{i(-\omega+i\frac{\Gamma_{\mu,\bf k}^+}{2})t_2}-1}{\big(-\omega-\Omega_{\mu,\bf k}^+-\mu+\frac{i\Gamma_{\mu,\bf k}^+}{2}\big)}-\frac{e^{i(\Omega_{\mu,\bf k}^--\mu)t_2}}{\Omega_{\mu,\bf k}^-}\frac{e^{i(-\omega+i\frac{\Gamma_{\mu,\bf k}^-}{2})t_1}+1}{\big(-\omega+\Omega_{\mu,\bf k}^--\mu+\frac{i\Gamma_{\mu,\bf k}^-}{2}\big)}\Bigg]\ .
\end{eqnarray}
The signs in the last equations are chosen so the reader can follow where the pole positions are. One can notice that the difference from equation (\ref{i1chemicalnonequi}) and (\ref{i2chemicalnonequi}) beside from the time $t_1$ and $t_2$ comes from the sign difference between the chemical potential and the new energies $\Omega_{\mu,\bf k}^{\pm}$, where the plus $+\mu$ only accompanies $\Omega_{\mu,\bf k}^{+}$ and $-\mu$ accompanies $\Omega_{\mu,\bf k}^{-}$. It is intuitive then to define: 
\begin{equation}
    \Omega_{\mu,\bf k}^{++}\equiv\Omega_{\mu,\bf k}^{+}+\mu\ ,
\end{equation}
and
\begin{equation}
    \Omega_{\mu,\bf k}^{--}\equiv\Omega_{\mu,\bf k}^{-}-\mu\ .
\end{equation}

Notice that the decay parameters $\Gamma_{\bf k}^{\pm}$ remain the same . As the inhomogeneous solution is the integral over $\omega$ of the multiplication of $I_1$ and $I_2$. The poles of (\ref{gmemnonequichem}) are given by a combination of 
\begin{equation}
    \omega=\pm \Omega_{\mu,\bf k}^{\pm\pm}\pm i\frac{\Gamma_{\mu,\bf k}^{\pm}}{2}\ .
\end{equation}

%=================================================================================================================
\section*{Solution for small chemical potential}
\label{s3}
%=================================================================================================================
IIn the limit where the chemical potential is much smaller than the particle energy, one can assume that the modified energies and decay parameters are of the same order. That is,
\begin{equation}
    \Omega_{\bf k}^{\pm}\rightarrow \Omega_{\bf k}\ ,\hspace{1cm}
    \Gamma_{\bf k}^{\pm}\rightarrow \Gamma_{\bf k}\ .
\end{equation}
This will give us the advantage to use the self energy $\Pi_{\bf}^+$ that appears in (\ref{secondKBequationfinitemu}) as a decay parameter $\Gamma_{\bf}$ after evaluated at a pole and using the KMS condition.  With these approximations we can write (\ref{gmemnonequichem}) :
\begin{eqnarray}
    G_{\mu,\bf{k}}^{mem}(t_1,t_2)=&&\frac{-i}{2\Omega_{\bf k}^3}\int\frac{d\omega}{2\pi}\frac{e^{-i\omega(t_1-t_2)}\coth\big(\frac{\beta \omega}{2}\big)\Pi^-_{\bf k}(\omega)}{\big((\omega+\mu+i\Gamma_{\bf k}/2)^2-\Omega_{\bf k}^2\big)(\big((\omega+\mu-i\Gamma_{\bf k}/2)^2-\Omega_{\bf k}^2\big)}\nonumber\\
   && \times\bigg[\big(i(\omega+\mu)\sin(\Omega_{\bf k}t_1)-\Omega_{\bf k}\cos(\Omega_{\bf k}t_1\big)e^{i(\omega+\mu+i\Gamma_{\bf k}/2)t_1}+\Omega_{\bf k}]\nonumber\\
   && \times\bigg[\big(i(\omega+\mu)\sin(\Omega_{\bf k}t_2)+\Omega_{\bf k}\cos(\Omega_{\bf k}t_2\big)e^{-i(\omega+\mu-i\Gamma_{\bf k}/2)t_1}-\Omega_{\bf k}]\ .
\end{eqnarray}
The expansion of every sine and cosine must be performed in order to effectively use the correct pole, we can choose to retain the chemical potential in the denominator or we can perform the change $\omega'=\omega+\mu$. As before, it is convenient to write the last equation using the notation
\begin{equation}
    \tilde S_i=\sin(\Omega_{\bf k}t_i)e^{i(\omega'+i\Gamma_{\bf k}/2)t_i}; \hspace{0.5cm}\tilde C_i=\cos(\Omega_{\bf k}t_i)e^{i(\omega'+i\Gamma_{\bf k}/2)t_i}\ ,
\end{equation}
in order to obtain:
\begin{eqnarray}
    G_{\mu,\bf{k}}^{mem}(t_1,t_2)=&&\frac{-i}{2\Omega_{\bf k}^3}\int\frac{d\omega'}{2\pi}\frac{e^{-i(\omega'-\mu)(t_1-t_2)}\coth\big(\frac{\beta (\omega'-\mu)}{2}\big)\Pi^-_{\bf k}(\omega'-\mu)}{\big((\omega'+i\Gamma_{\bf k}/2)^2-\Omega_{\bf k}^2\big)(\big((\omega'-i\Gamma_{\bf k}/2)^2-\Omega_{\bf k}^2\big)}\nonumber\\
    &&\times\bigg[-\omega'^2\tilde S_1\tilde S_2^*+i\omega'\Omega_{\bf k}\tilde S_1\tilde C_2^*-i\omega'\Omega_{\bf k}\tilde C_1\tilde S_2^*\nonumber\nonumber\\
    &&\hspace{0.8cm}+i\omega'\Omega_{\bf k}\tilde S_1+\Omega_{\bf k}^2\tilde C_1+i\omega'\Omega_{\bf k}\tilde S_2^*+\Omega_{\bf k}^2\tilde C_2^*-\Omega_{\bf k}^2\bigg]\ .
\end{eqnarray}
The lengthy calculation can be easily performed by taking into account convergence at each time. The definition of the decay parameter still holds under this approximation, i.e., $\Gamma^{\pm}_{\bf k}(\Omega_{\bf k})\approx-\text{Im}\Pi^R(\Omega_{\bf k}\pm\mu)/\Omega_{\bf k}\approx\Gamma_{\bf k}$. Remembering that the symmetry of the involved functions must be performed as $t_1\rightarrow t_2$ then one needs to apply $\omega'\rightarrow -\omega'$ and $\mu\rightarrow -\mu$. We finally arrive to
\begin{eqnarray}\label{memorychemicalpotential}
     G_{\mu,\bf{k}}^{mem}(t_1,t_2)=&&\frac{1}{2\Omega_{\bf k}}\Bigg(\coth\bigg(\frac{\beta(\Omega_{\bf k}-\mu)}{2}\bigg)+\coth\bigg(\frac{\beta(\Omega_{\bf k}+\mu)}{2}\bigg)\Bigg)\nonumber\\
     &&\times\cos(\Omega_{\bf k}(t_1-t_2))\Bigg[e^{-\frac{\Gamma_{\bf k}}{2}(t_1+t_2)}-e^{-\frac{\Gamma_{\bf k}}{2}(t_1-t_2)}\Bigg]e^{i\mu(t_1-t_2)}\ .
\end{eqnarray}
Under this approximation the factor $e^{i\mu(t_1-t_2)}$ is still around, and we obtain a somehow expected dependence of the chemical potential in the hyperbolic cotangents. 

The homogeneous solution $\hat{G}_{\mu,\bf k}^{+}(t_1,t_2)$ of (\ref{secondKBequationfinitemu}) depends on the initial condition of the statistical propagator, and is given by a generalization of (\ref{initialcoditionsdeltaplus})
\begin{eqnarray}
    \hat{G}_{\mu,\bf k}^{+}(t_1,t_2)=&&G_{in}^+\dot{G}_{\mu,\bf k}^-(t_1)\dot{G}_{\mu,\bf k}^-(t_2)+\ddot{G}_{in}^+G_{\mu,\bf k}^-(t_1)G_{\mu,\bf k}^-(t_2)\nonumber\\
    &+&\dot{G}_{in}^+\big(\dot{G}_{\mu,bf k}^-(t_1)G_{\mu,\bf k}^-(t_2)+G_{\mu,\bf k}^-(t_1)\dot{G}_{\mu,\bf k}^-(t_2)\big)\ .
\end{eqnarray}
As before we can choose a no particle initial condition, in which we have $G^+_{in}=1/\Omega_{\bf k}$, $\dot G^+_{in}=0$  and $\ddot G^+_{in}=\Omega_{\bf k}$, and using the spectral function given in (\ref{spectralfunctminusplusapprox}) we have
\begin{eqnarray}
       \hat{G}_{\mu,\bf k}^{+}(t_1,t_2)=&&\frac{1}{\Omega_{\bf k}}\big[\Omega_{\bf k}\cos(\Omega_{\bf k}t_1)+i\mu\sin(\Omega_{\bf k}t_1)\big]e^{i(\mu+i\Gamma_{\bf k}/2)t_1}\nonumber\\
       &\times&\big[\Omega_{\bf k}\cos(\Omega_{\bf k}t_2)-i\mu\sin(\Omega_{\bf k}t_2)\big]e^{-i(\mu-i\Gamma_{\bf k}/2)t_2}\nonumber\\
       &+&\frac{1}{\Omega_{\bf k}}\sin(\Omega_{\bf k}t_1)\sin(\Omega_{\bf k}t_2)e^{i\mu(t_1-t_2)}e^{-\frac{\Gamma_{\bf k}}{2}(t_1+t_2)}\ .
\end{eqnarray}
Under the approximation where $\mu\ll\Omega_{\bf k}$, we can write the above equation as
\begin{equation}\label{homogeneousinitialchemicalpotential}
     \hat{G}_{\mu,\bf k}^{+}(t_1,t_2)=\frac{1}{\Omega_{\bf k}}\cos(\Omega_{\bf k})e^{i\mu(t_1-t_2)}e^{-\frac{\Gamma_{\bf k}}{2}(t_1+t_2)\ .}
\end{equation}
We finally obtain the full statistical propagator for small chemical potential by adding (\ref{memorychemicalpotential}) and (\ref{homogeneousinitialchemicalpotential}) in order to obtain
\begin{eqnarray}
    G^+_{\mu,\bf k}(t_1,t_2)=&&\frac{\cos\big(\Omega_{\bf k}(t_1-t_2)\big)e^{i\mu(t_1-t_2)}}{2\Omega_{\bf k}}\Bigg[\bigg(f_{eq}(\Omega_{\bf k}+\mu)+f_{eq}(\Omega_{\bf k}-\mu)\bigg)e^{-\frac{\Gamma_{\bf k}}{2}(t_1+t_2)}\nonumber\\
   &&+\frac{1}{2}\Bigg(\coth\bigg(\frac{\beta(\Omega_{\bf k}+\mu)}{2}\bigg)+\coth\bigg(\frac{\beta(\Omega_{\bf k}-\mu)}{2}\bigg)\Bigg)e^{-\frac{\Gamma_{\bf k}}{2}|t_1-t_2|}\Bigg]\ .
\end{eqnarray}
As for the case of the spectral function, the chemical potential will have a $\pm \mu$ contribution at first order in the statistical propagator. For a fixed $\Gamma_{\bf k}|t_1-t_2|$ we can take the limit $\Gamma_{\bf k}(t_1+t_2)\rightarrow\infty$, that will lead to an equilibrium condition and regain the already computed equilibrium statistical propagator.
%=================================================================================================================
 \section{Conclusions}
\label{s4}
%=================================================================================================================
In this work, we have explored the structure and dynamics of bosonic propagators within the real-time formalism of quantum field theory, focusing particularly on systems at finite temperature and chemical potential. Using the Schwinger-Keldysh contour, we systematically constructed the contour-ordered Green’s functions and identified their physical components.

In out-of-equilibrium systems, the Kubo-Martin-Schwinger (KMS) condition no longer holds, necessitating a fully dynamical treatment of the two-point functions. To address this, we derived solutions to the Kadanoff-Baym equations for both the spectral function and the statistical propagator. These propagators provide a robust framework for describing the system's dynamics.

Our analysis demonstrates the versatility of the real-time formalism in incorporating chemical potential. We investigated both equilibrium and non-equilibrium phenomena in bosonic field theories, with particular attention to how the chemical potential affects boundary conditions, introduces new poles, modifies decay parameters, and alters distribution functions. The methods and results presented here lay the groundwork for calculating particle production rates, response functions, and transport coefficients across a wide range of physical systems—from scalar field dynamics in the early universe to excitations in hot and dense matter created in heavy-ion collisions.

Future directions include incorporating the chemical potential at higher orders in the pole structure and self-energy in a self-consistent manner, extending the analysis to gauge fields, and studying symmetry breaking and phase transitions using non-equilibrium Green’s function techniques. These developments will further deepen our understanding of quantum fields in complex thermal environments and dynamically evolving media.

%=================================================================================================================
\bibliography{biblio}% Produces bibliography via BibTeX.
%=================================================================================================================

\end{document}